\documentstyle[12pt,aasms4]{article}

\received{: 17 December 1996}
\accepted{: 29 January 1997}

\slugcomment{To appear in the Astrophysical Journal Letters}

\lefthead{SENGUPTA}
\righthead{MAGNETIC FIELD DECAY IN NEUTRON STARS}
\begin{document}

\title{GENERAL RELATIVISTIC EFFECTS ON THE OHMIC DECAY OF CRUSTAL MAGNETIC
FIELDS IN NEUTRON~STARS}

\author{SUJAN SENGUPTA}
\affil{Indian Institute of Astrophysics \\ Koramangala, Bangalore 560 034,
India \\ sujan@iiap.ernet.in}

\begin{abstract}

The ohmic decay of magnetic fields confined within the crust of neutron stars
is considered by incorporating the effect of space-time curvature produced by the
intense gravitational field of the star. It is shown that general relativistic
effect reduces the magnetic field decay rate substaintially and specially at
the late time of the evolution the decay rate decreases by several orders of
magnitude when compared with the case without the relativistic effect.

\end{abstract}


\section{INTRODUCTION}

The magnetic field evolution in neutron stars has been a subject of much
discussion over the years both in the observational and in the theoretical
context. Calculations of ohmic decay of dipolar magnetic fields were performed
by Sang \& Chanmugam (1987) who demonstrated that the field does not decay
exponentially. The reviews by Lamb (1991), Chanmugam (1992) and Phinney \&
Kulkarni (1994) provide the present understanding on the decay of magnetic
fields in isolated neutron stars. Haensel, Urpin \& Yakovlev (1990) pointed
out to the possibility that magnetic fields in the core could decay rapidly
by ambipolar diffusion. The studies of magnetic field configurations in which
the field vanishes in the stellar core (\cite{SC87}; \cite{CS89}; \cite{UM92};
\cite{UV93}) show that because of the relatively low electrical conductivity
of crustal matter, the decay times may be short enough to be of observational
interest if the impurity concentration is high and the field is initially
confined to a small part of the crust. 
On the other hand, considering field configurations that do not vanish in the
core, Pethick \& Sahrling (1995) showed that the shortest possible decay time
is about two orders of magnitude longer than the characteristic time-scale
for decay of configurations in which the magnetic field vanishes in the core.

In all the investigations, an important feature of neutron stars, the space-time
curvature produced by the intense gravitational field has not been taken into
account although it is well-known that the space-time curvature exterior and
interior to neutron stars can significantly alter the electromagnetic field
(\cite{WS83}; \cite{S95}; \cite{BO95}).

In this letter it is demonstrated, by adopting a simplified model, that the
decay rate decreases significantly when the general relativistic effect is
taken into consideration.
It is, however, worth mentioning that due to lack of proper understanding
on the initial configuration of the magnetic field and of the impurity
parameter which plays crucial role in determining the electrical conductivity,
no model presented so far can be attributed to represent the quantitative
feature of the actual situation. Nevertheless, all these investigations
 are important from the
qualitative point of view and provide significant insight on the decay of
the magnetic field. The scope of the present work although is also
limited to an idealistic situation but it is sufficient to demonstrate the
important role played by curved space-time at the crust of neutron stars in
governing the decay of magnetic fields. The present result indicates that
the space-time curvature produced by the intense gravitational field at
the crust can lead to characteristic decay-times much longer than the
existing estimations.

\section{MAGNETIC FIELD EVOLUTION IN CURVED SPACE-TIME}

If hydrodynamic motions are negligible and the anisotropy of the electrical
conductivity is small, the induction equation in flat space-time can be
written as:
  \begin{equation}\label{FLAT}
\frac{\partial{\bf B}}{\partial t}=-\nabla \times \left(\frac{c^2}{4\pi\sigma}
\nabla \times B\right),
\end{equation}
where $\sigma$ is the electrical conductivity.

If a stationary gravitational field is taken into account, the corresponding 
induction equation in curved space-time can be written as:
\begin{equation}\label{GTR1}
\frac{\partial F_{kj}}{c\partial t}=\frac{c}{4\pi}
\left[\frac{\partial}{\partial x^k}
\left\{\frac{1}{\sqrt{-g}}\frac{1}{\sigma u^0}g_{ij}
\frac{\partial}{\partial x^l}(\sqrt{-g}F^{il})\right\}-
\frac{\partial}{\partial x^j}\left\{\frac{1}{\sqrt{-g}}
\frac{1}{\sigma u^0}g_{ik}\frac{\partial}{\partial x^l}(\sqrt{-g}
F^{il})\right\}\right],
\end{equation}
where $F_{ij}$ is the electromagnetic
field tensor, $g_{ij}$ is the components of space-time metric that describes the
background geometry, $g={\rm det}|g_{ij}|$ and $u^{0}$ is the time component of
the velocity vector of the fluid (I have considered $u^i=0$). 
Here and afterwards latin indices run over spatial co-ordinates only whereas
greek indices run over both time and space co-ordinates.

Now we need to choose a space-time metric which describes the geometry of the
region under consideration. Although, in Newtonian theory the gravitational
field of a rotating body is the same as that of a non-rotating body, 
in general relativity rotation affects the space-time geometry. However, in the
absence of any suitable metric that can describe the space-time geometry inside
a rotating neutron star, I consider, in the present investigation, a stationary 
and static gravitational field. Further, it is well-known (see for example
\cite{DTB95} and references therein) that the gravitational mass of the entire crust of neutron stars
consists of less than 3\% of the total mass of the star for any equation of
state and hence the self gravitation of the crust is negligible compared to the
gravitational field due to the core. Since in the present
work, the magnetic field is considered to be confined within the outermost
crust and the magnetic field strength therein is supposed to be sufficiently
low compared to the gravitational field so that the space-time curvature is not
affected by the electromagnetic field, one can very well adopt the exterior
Schwarzschild metric which on the other hand simplifies the calculations.
The metric is given by:
\begin{equation}
ds^2=(1-\frac{2m}{r})c^2dt^2-(1-\frac{2m}{r})^{-1}dr^2-r^2(d\theta^2+
\sin^2\theta d\phi^2),
\end{equation}
where $m=MG/c^2$, $M$ being the total gravitational mass of the core. Since,
the crust consists of less than a few percent of the total gravitational mass,
$M$ can be regarded as the total mass of the star. 

If $F_{(\alpha\beta)}$ are the components of the electromagnetic field tensor in
a local Lorentz frame, then the components of the electromagnetic field tensor
$F_{\gamma\delta}$ are defined in the curved space-time through the relation:
\begin{equation}\label{TET}
F_{(\alpha\beta)}=\lambda^{\gamma}_{(\alpha)}\lambda^{\delta}_{(\beta)}
F_{\gamma\delta},
\end{equation}
where $\lambda^{\alpha}_{(\beta)}$ are the non-zero components of the orthonormal
tetrad of the local Lorentz frame for the Schwarzschild geometry given in
Sengupta (1995).

Following the convention, I restrict myself by the consideration of the decay of
a dipolar magnetic field which has axial symmetry so that the vector potential
${\bf A}$ may be written as $(0,0,A_{\phi})$ in spherical polar co-ordinates
 where $A_{\phi}=A(r,\theta,t).$ Since the hydrodynamic motion is negligible
 so $u^i= dx^i/ds = 0$ and the metric gives
$u^0=(1-2m/r)^{-1/2}.$

Substituting the metric components in equation (\ref{GTR1}) and
using the definition $F_{\alpha\beta}=(A_{\beta,\alpha}-A_{\alpha,\beta})$, 
 one obtains the induction equation in Schwarzschild geometry in term of
the vector potential as:
\begin{equation}\label{GTR2}
\frac{\partial A_{\phi}}{\partial t}=\frac{c^2}{4\pi\sigma}(1-
\frac{2m}{r})^{1/2}\sin\theta\left[\frac{\partial}{\partial r}\left\{(1-
\frac{2m}{r})\frac{1}{\sin\theta}\frac{\partial A_{\phi}}{\partial r}
\right\}+\frac{\partial}{\partial \theta}\left(\frac{1}{r^2\sin\theta}
\frac{\partial A_{\phi}}{\partial \theta}\right)\right].
\end{equation}

Now for the flat space-time I choose $A_{\phi}=f(r,t)\sin\theta/r$
where $r$ and $\theta$ are the spherical radius and polar angle
respectively and one gets from equation (\ref{FLAT})
\begin{equation}\label{FLAT1}
\frac{\partial^2f(x,t)}{\partial x^2}-\frac{2}{x^2}f(x,t)=\frac{4\pi R^2
\sigma}{c^2}\frac{\partial f(x,t)}{\partial t},
\end{equation}
where $x=r/R$ and $R$ is the radius of the star.

For the general relativistic case, the choice is guided by the form of the
time-independent dipole magnetic field in Schwarzschild geometry obtained
by Wasserman and Shapiro (1983) and can be written as
$A_{\phi}=-g(x,t)\sin^2\theta$.
Hence, from equation (\ref{GTR2}) one obtains
\begin{equation}\label{GTR3}
(1-\frac{y}{x})^{1/2}\left[(1-\frac{y}{x})\frac{\partial^2g(x,t)}{\partial x^2}
+\frac{y}{x^2}\frac{\partial g(x,t)}{\partial x}-\frac{2}{x^2}g(x,t)\right]=
\frac{4\pi R^2\sigma}{c^2}\frac{\partial g(x,t)}{\partial t},
\end{equation}
where $y=2m/R$

For both the cases, I impose the usual boundary conditions which are (i)
in the deep layer of the crust, the magnetic field vanishes and (ii) at the
outer boundary of the crust, the field matches onto an exterior dipole magnetic
field.

\section {THE MODEL}

I shall follow Urpin \& Van Riper (1993) in my approach with one exception.
To simplify the calculations, I neglect the neutron star cooling which can
significantly decrease the field decay rate since the conductivity of the
crust depends on the temperature $T$. The qualitative nature of the present
results, however, will not be altered if one incorporates the neutron star
cooling as long as $\sigma$ itself is independent of the space-time curvature.
Calculations of ohmic decay of bipolar magnetic field without the effect
of the neutron star cooling was performed by Sang \& Chanmugam (1987). 
Their results have been used to check the numerical accuracy of the present
results corresponding to flat space-time.

As mentioned in the introduction, any given magnetic field configuration
in flat space-time would be modified by the curvature of space-time produced
by the gravitational field. If initially (at $t=0$)
$$ A_{\phi}(r,\theta,0)=A_{\phi}(r,\theta)=\frac{f(r,0)}{r}\sin\theta=
\frac{f(r)}{r}\sin\theta$$
 for flat space-time, then for curved space-time
$$ A_{\phi}(r,\theta,0)=A_{\phi}(r,\theta)=-\frac{3rf(r)}{8m^3}s(r)
\sin^2\theta ,$$
where $s(r)$ is the general relativistic correction factor. Hence, it
can be shown (\cite{WS83}) that if one assumes the initial value of
 $f(r,t)=f(r)$ for flat space-time, then for curved space-time 
  \begin{equation}
g(r,0)=g(r)=\frac{3rf(r)}{8m^3}[r^2\ln(1-\frac{2m}{r})+2mr+2m^2].
  \end{equation}
Clearly, if $r\rightarrow\infty, \; g(r)\rightarrow f(r).$

I have considered the decay of the magnetic field which initially occupies
the surface layers of the crust upto a depth $x=0.955$ which corresponds to the
density $5\times 10^{11}$ $\rm{gcm^{-3}}$.

Following the approaches of Urpin and Van Riper (1993) I have calculated
the electrical conductivity within the crust that has been derived by Urpin \&
Yakovlev (1980). The effect of electron-ion scattering has been neglected
since the region where this effect could be important is sufficiently thin.
The impurity parameter $Q$ has been taken as $0.001$ and the conductivity
is calculated by assuming the region under consideration to be isothermal
with a constant temperature $T=10^7 \; K$. Results with $T=10^5 \; K$
have also been presented for few relevant cases.  The calculations have been performed
by considering a neutron star of mass $1.4M_{\odot}$ and radius 10.6 km.

\section{RESULTS AND DISCUSSIONS}

Equations (\ref{FLAT1}) and (\ref{GTR3}), with the corresponding boundary
conditions, have been solved numerically. The calculations have been performed
by making use of the standard Crank-Nicholson differencing scheme. 
 For the sake of comparison of the results, the magnetic field in
Schwarzschild geometry has been transformed into a local Lorentz frame by
using equation (\ref{TET}). If one takes the mass $M$ sufficiently low
($< 0.01M_{\odot}$) then the decay profile calculated in the local Lorentz
frame coincides with that for the flat space-time and for the present purpose
this provides sufficient check for the numerical accuracy of the results.
The effect of general relativity in governing
the magnetic field decay is clearly depicted in the figures. Figure 1 shows
the evolution of the surface magnetic field normalized to its initial value
for both the general relativistic and non-relativistic cases. The 
qualitative nature of the decay profiles are same with that presented by
Sang \& Chanmugam (1987) as in both the cases the effect of the neutron star
cooling has not been taken into consideration. In the present work such type
of effect is not relevant since it is expected that the cooling process and
hence the electrical conductivity are independent of the space-time curvature.
\placefigure{fig1}

In Figure 1 the decay of the surface magnetic field is presented for neutron
stars with masses $1M_{\odot}$, $1.4M_{\odot}$ and $1.8M_{\odot}$ and with
the same radius $R=10.6$ km. However, the electrical conductivity is calculated
by adopting $1.4M_{\odot}$ mass configuration. It is clear from the figure that
the decay rate decreases significantly with the increase in mass, i.e., with
the increase in the gravitational field for the whole period of evolution.
Unlike the case for flat space-time, the decay rate gradually slows down as
the time increases. In other words, the effect of general relativity becomes
more significant with the increase in the age of the star. At the late stage
of evolution, say, after 3 Gyr, the difference in the strength of the surface magnetic field
between the two cases becomes as large as about 3 orders of magnitude. It is 
interesting to note that if the mass of the star is as high as $1.8M_{\odot}$
which can be possible with any stiff equation of state of matter inside the 
star, then due to the effect of space-time curvature the decay rate
decreases so dramatically that when the other physical effects such as cooling
of the neutron star which too play significant role in reducing the decay rate,
are incorporated the decay in the magnetic
field could be too small to be determined observationally and during the
whole life span of the star the strength of the magnetic field would remain
 almost the same with that of its initial value. Since general relativistic
effect alone can reduce the decay rate  significantly, observational
evidence of faster decay rate may constrain the compactness of the neutron
star and hence the equation of state of matter inside the star. It is worth
mentioning here that the general relativistic effect on the magnetic field
is less uncertain than any other physical phenomenon exterior or interior to
isolated neutron stars.
\placefigure{fig2}

In Figure 2 the variation of the magnetic field along the radial
points inside the crust at different times are presented for a neutron star
of mass $1.4M_{\odot}$. The initial
field strength increases by the inclusion of general relativistic effect.
This result is well known for time-independent dipole magnetic field 
(\cite{WS83}; \cite{S95}). After 5 Myr the strength of the magnetic field
for the curved space-time becomes almost equal to the initial value of the 
field strength when the general relativistic effect is not included. 
Subsequently, for flat space-time the field strength at any time remains
much lower  than that for curved space-time. However, it cannot be inferred
that due to the modification in the initial configuration of the magnetic
field the final field strength becomes much higher for the general
relativistic case than that for the case of flat space-time. Rather, the entire
evolution should be modified by the general relativistic effect as indicated
by equation (\ref{GTR3}). As the temperature of the crust must decrease 
with time, the electrical conductivity becomes higher and hence at the late stage
of evolution the decay rate should decrease substaintially irrespective of
the nature of the surrounding space-time. In figure 2,
the magnetic field strength at 1 Gyr with the crust temperature $T=10^5$ K
for both the general relativistic and non-relativistic cases are presented.
 The results provide a confirmation of the above fact. Since the magnetic
field strength for the general relativistic case is almost double to its value
when relativistic effect is not considered, it is very much likely that the
magnetic field decay should be too small to be detected observationally if
both the cooling effect and the general relativistic effect are taken
together into account. Therefore,
detail investigations by considering the neutron star cooling and other
physical phenomenon alongwith the general relativistic effect will be of
much theoretical interest.

\section{CONCLUSIONS}

The important message which is conveyed by the present calculations is that
whatever be the electrical conductivity of the crustal material, high or low,
irrespective of the question whether the magnetic field vanishes at the core or not and
whatever be the impurity content of the neutron star crust, the decay time
of the magnetic field is lengthened by the intense gravitational field that
certainly exists inside the star. Sang \& Chanmugam (1987) showed that
the decay is not exponential while Urpin \& Muslimov (1992) pointed out 
that even if the magnetic field is initially absent in the core,
diffusion of the field into a highly conducting core would retard the
surface field decay. Pethic and Sahrling (1995) suggested
 that if long decay times were established
observationally, these could not necessarily imply as the evidence for matter
in the stellar core having a high conductivity. Irrespective of all the
uncertainties that still exist and require further theoretical investigations
as well as observational evidences the present result can atleast provide a
firm understanding that general relativity is certainly responsible if the
decay time is indeed very long. Therefore, the present demonstration is
important in the sense that it establishes a concrete restrictions on the
theoretically possible ways of obtaining short ohmic decay times for magnetic
fields in neutron stars.  Further, the present results provide an interesting
feature that the more is the compactness of the neutron star the longer
is the decay times and hence the general relativistic effects on the decay
of magnetic field could be a possible tool for restricting the equation
of state of matter inside the star which determines the compactness of
the neutron star.

The crucial lesson that the present results provide is that, in addition to
the calculations of the conductivity with better estimation of the impurity
content, the incorporation of the effect of superfluidity and superconductivity 
in the core and other physical effects such as Hall drift, full consideration
of the general relativistic effects must be given in order to
make more realistic estimates of decay times of the magnetic field.

\acknowledgments

I am grateful to late Professor N. C. Rana (1954-1996) who initiated this
work and had been a constant source of encouragement till his sudden demise. 
Thanks are due to the anonymous referee for useful comments and 
constructive criticisms.

\clearpage
\plotone{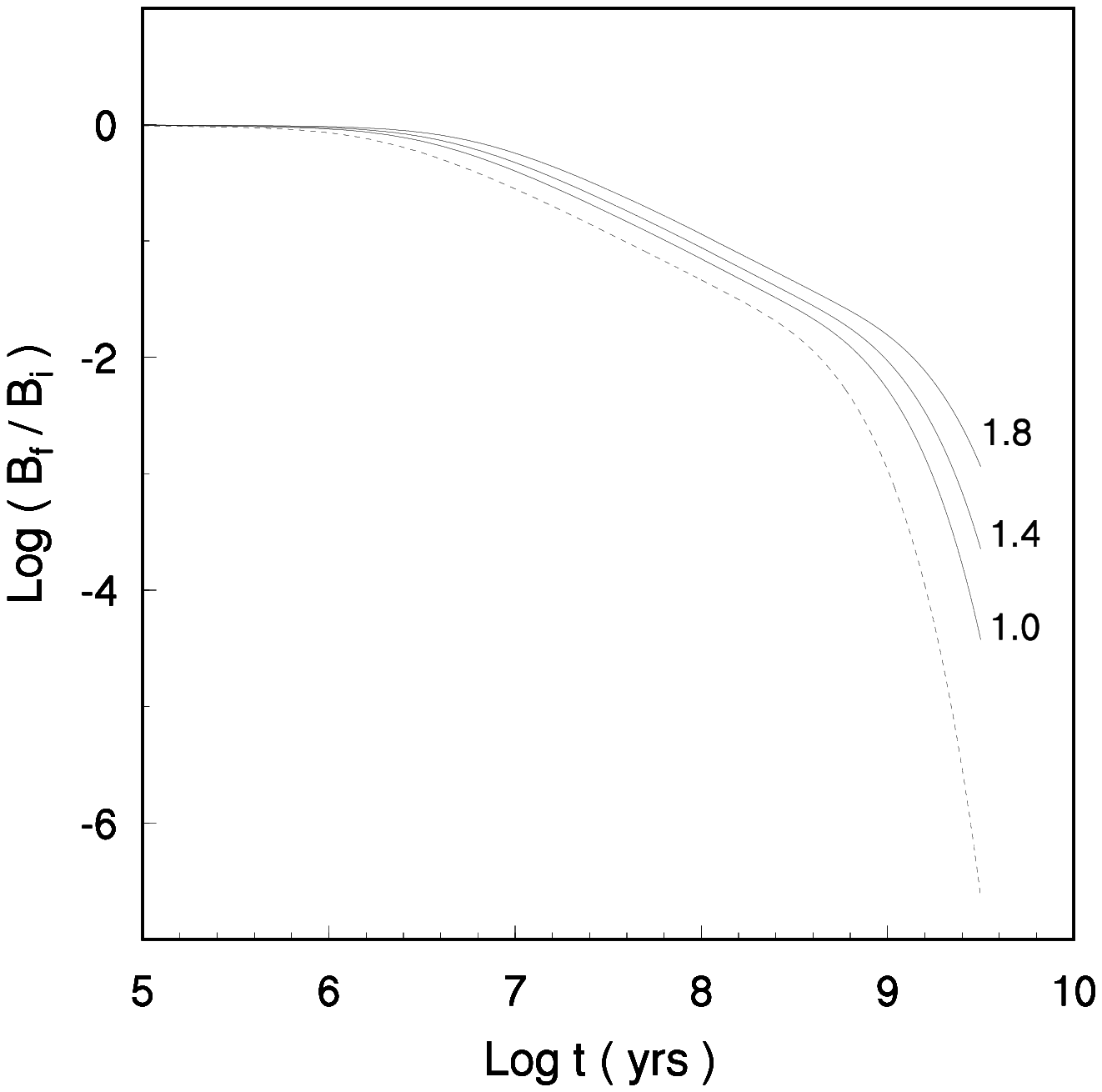}
\figcaption[figure1.eps]{The evolution of surface magnetic field normalized
to its initial value for flat and curved spacetimes. Solid lines represent
the results for curved space-time while broken lines represent that for flat
space-time. The numbers near the curves indicate mass in $1 \; M_{\odot}$.
For all the cases the radius is taken as $R=10.6$ km and the electrical
conductivity is calculated for a $1.4 \; M_{\odot}$ neutron star crust.
\label{fig1}}

\plotone{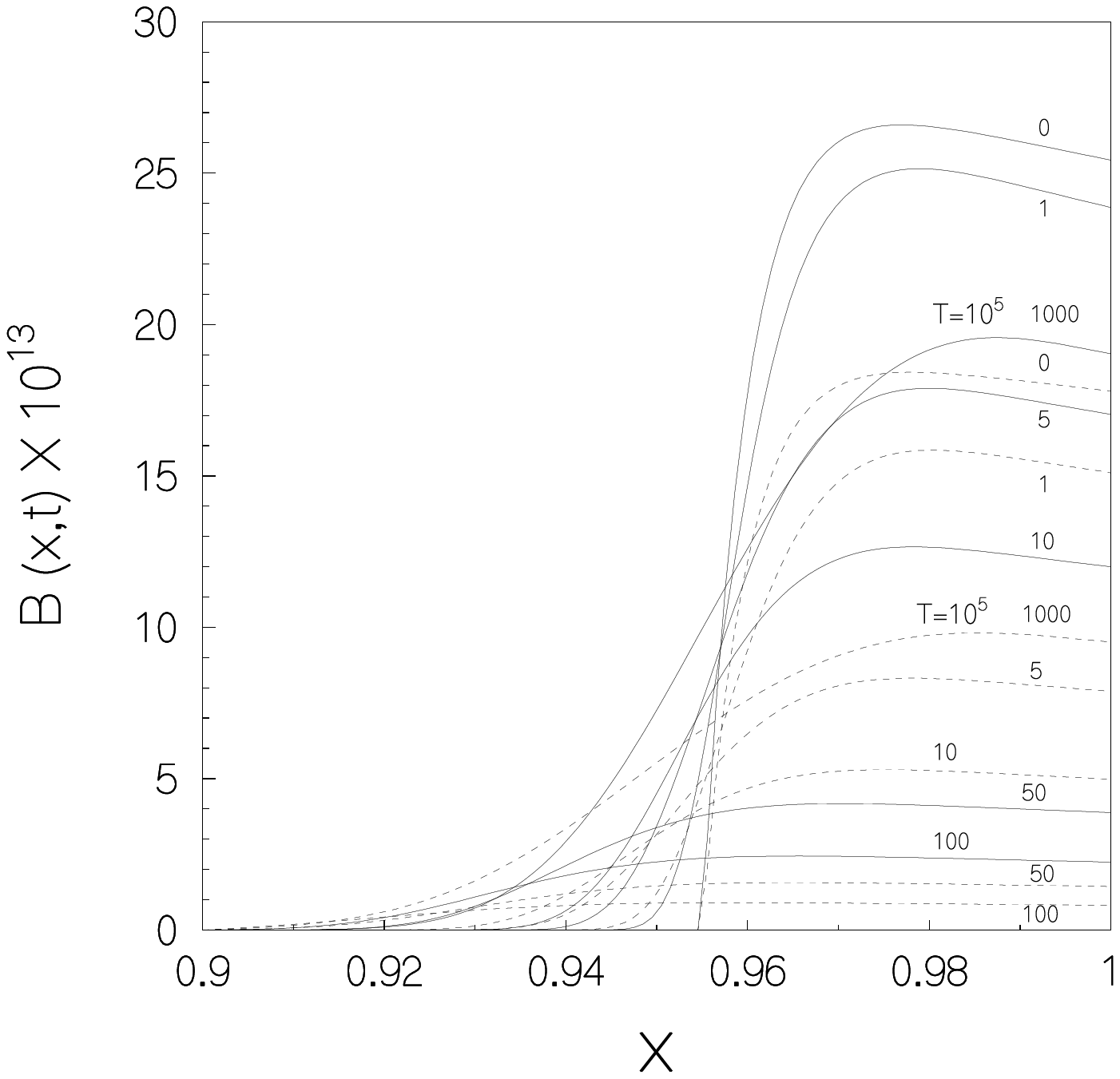}
\figcaption[figure2.eps]{The dipole magnetic field at different time
 along the normalized radial points. The numbers near the curves indicate
 $t$ in Myr and the curves corresponding to $t=0$ indicate the initial
value of the magnetic field. Unless indicated, all the curves present the
results with temperature $T=10^7$ K. Solid lines represent the case for curved
space-time while broken lines represent that for flat space-time. \label{fig2}}

\end{document}